\documentclass[twocolumn,floatfix, nofootinbib,prd,preprintnumbers,showpacs,showkeys,superscriptaddress]{revtex4-1}

\usepackage{times} 
\usepackage{amsfonts}
\usepackage{amsmath}
\usepackage{amssymb}
\usepackage{booktabs}
\usepackage{braket}
\usepackage{color}
\usepackage{graphicx}
\usepackage{hyperref}
\usepackage{cleveref}
\usepackage{mathtools}
\usepackage[rightcaption]{sidecap}
\usepackage{soul}
\usepackage{subfigure}
\usepackage{tikz}
\usetikzlibrary{calc, decorations.markings, decorations.pathmorphing}

\definecolor{amaranth}{rgb}{0.9, 0.17, 0.31}
\definecolor{brightpink}{rgb}{1.0, 0.0, 0.5}
\definecolor{cornflowerblue}{rgb}{0.39, 0.58, 0.93}
\definecolor{deepcarminepink}{rgb}{0.94, 0.19, 0.22}
\definecolor{palatinateblue}{rgb}{0.15, 0.23, 0.89}
\definecolor{radicalred}{rgb}{1.0, 0.21, 0.37}
\definecolor{vividviolet}{rgb}{0.62, 0.0, 1.0}

\hypersetup{linktoc=all,
    colorlinks, linkcolor={palatinateblue},
    citecolor={brightpink}, urlcolor={amaranth}
}

\newcommand{\be}{\begin{equation}}
\newcommand{\ee}{\end{equation}}
\newcommand{\bes}{\begin{subequations}}
\newcommand{\ees}{\end{subequations}}
\newcommand{\bea}{\begin{eqnarray}}
\newcommand{\eea}{\end{eqnarray}}

\begin{document}

\title{On the duality of Schwarzschild-de Sitter spacetime and moving mirror}

\author{Diego Fernández-Silvestre}
\email{diefer2@alumni.uv.es}
\affiliation{Departamento de Física Teórica and IFIC, Universidad de Valencia-CSIC, C. Dr. Moliner 50, 46100 Burjassot, Spain.}

\author{Joshua Foo}
\email{joshua.foo@uqconnect.edu.au}
\affiliation{Centre for Quantum Computation \& Communication Technology, School of Mathematics \& Physics,
University of Queensland, St.~Lucia, Queensland, 4072, Australia.}

\author{Michael R.R. Good}
\email{michael.good@nu.edu.kz}
\affiliation{Department of Physics \& Energetic Cosmos Laboratory, Nazarbayev University,
Kabanbay Batyr Ave 53, Nur-Sultan, 010000, Kazakhstan.}

\begin{abstract}
The Schwarzschild-de Sitter (SdS) metric is the simplest spacetime solution in general relativity with both a black hole event horizon and a cosmological event horizon. Since the Schwarzschild metric is the most simple solution of Einstein's equations with spherical symmetry and the de Sitter metric is the most simple solution of Einstein's equations with a positive cosmological constant, the combination in the SdS metric defines an appropriate background geometry for semi-classical investigation of Hawking radiation with respect to past and future horizons. Generally, the black hole temperature is larger than that of the cosmological horizon, so there is heat flow from the smaller black hole horizon to the larger cosmological horizon, despite questions concerning the definition of the relative temperature of the black hole without a measurement by an observer sitting in an asymptotically flat spacetime. Here we investigate the accelerating boundary correspondence (ABC) of the radiation in SdS spacetime without such a problem. We have solved for the boundary dynamics, energy flux and asymptotic particle spectrum. The distribution of particles is globally non-thermal while asymptotically the radiation reaches equilibrium. 
\end{abstract}



\maketitle

\section{Introduction}
Historically, black holes have been important systems for investigating and understanding our universe. Theoretically, they have played a prominent role in attempts at unifying quantum theory and gravity. Along astrophysical lines of research, investigations into their formation and evolution have contributed to our knowledge of the large-scale structure of galaxies, while the detection of gravitational lensing and gravitational waves have corroborated the predictions of general relativity. More largely, the study of cosmological spacetimes is of great interest because we live in an accelerated expanding universe \cite{SupernovaSearchTeam:1998fmf, SupernovaCosmologyProject:1998vns}. The simplest expanding spacetime is de Sitter, which is especially appropriate to describe the early phase of exponential expansion predicted by inflation \cite{Guth:1980zm}. Then, the study of cosmological black holes (e.g. asymptotically de Sitter black holes) is well-motivated, for example, to investigate black holes formed during the early phase of inflation or even the global structure of black holes in the current phase of accelerating expansion.

To theoretically analyze black holes embedded in an expanding universe, the simplest spacetime to start is Schwarzschild-de Sitter spacetime (SdS), which is an exact black hole solution to the Einstein's equations with a positive cosmological constant. In the context of quantum field theory, the SdS background is particularly convenient to study the effects of event horizons on the propagation of quantum fields, since there are two different horizons in a single spacetime metric: a black hole horizon and a cosmological horizon.

In the asymptotically flat Schwarzschild black hole case, Hawking discovered the quantum particle creation effects that occur when a black hole forms from the gravitational collapse of a star \cite{Hawking:1974sw} (see also \cite{Unruh:1976db}). The particles are produced at late times in a thermal spectrum (with a grey-body factor) at a temperature that is proportional to the surface gravity at the event horizon of the black hole. It was further shown in analogy with the Schwarzschild black hole result, that the de Sitter cosmology also involves thermal particle production with temperature proportional to the surface gravity at the cosmological event horizon \cite{GibbonsHawking}. 

In the asymptotically de Sitter Schwarzschild black hole case, the particle production and thermodynamics were first investigated in \cite{GibbonsHawking}. One can demonstrate, using the mechanism of \cite{Hawking:1974sw}, that the black hole horizon of SdS spacetime, and the cosmological horizon of SdS spacetime, will create particles in a Planck distribution (with a grey-body factor) at a temperature that is proportional to the surface gravity of the corresponding horizon of SdS spacetime \cite{Bhattacharya:2018ltm}. The grey-body effect caused by the effective potential is an important physical aspect of the radiation. However, in the geometric optics approximation (i.e. no grey-body effect), there is only particle production in a thermal spectrum for the black hole at late times (after the collapse). Analogously, at early times (before the collapse), there is only cosmological particle production in a thermal spectrum. Our investigation, in part, is to compute and analyze the interesting interaction behaviour between both early-time and late-time thermal spectra. 

The above mentioned results of SdS spacetime came from considering that the particle detection is made by accelerating (but static) observers not at but close to the (future) cosmological horizon \cite{Bhattacharya:2018ltm}. It is more natural to study the quantum particle creation effects with respect to inertial observers instead of non-inertial observers. However, particle production in SdS spacetime and detection by inertial observers can be understood from detection by non-inertial (who are static) observers through amplitudes defined via the Bogoliubov coefficients \cite{Qiu:2019qgp}. In this sense, it is simpler to assume the perspective of accelerating observers in the analysis of particle production effects in SdS spacetime. Regardless of this question, there may be a gravitational analog model that has a well-defined and intuitive asymptotically flat spacetime inertial observer ready to detect the corresponding radiation.  This analog system is the moving mirror model.

The analogy of moving mirrors to black holes \cite{DeWitt:1975ys}, being a (1+1)-dimensional case of the dynamical Casimir effect \cite{moore1970quantum}, was initially introduced by Fulling and Davies \cite{Davies:1976hi, Davies:1977yv}. They determined that an accelerated boundary (moving mirror) with the appropriate trajectory in Minkowski spacetime creates particles in a thermal spectrum (at late times), analogously to Hawking's particle creation by black holes result \cite{Hawking:1974sw}. A standard approach to the mirror model is to assume a trajectory that fulfills given physical requirements and then calculate the corresponding radiation and resulting spectrum \cite{walker1985particle,carlitz1987reflections}. A central theoretical achievement of the mirror model has been the demonstration that these accelerated point mirrors non-trivially disturb the quantum vacuum via a non-zero Bogoliubov coefficients (and renormalized stress tensor), resulting in particle production, energy flux, entanglement and other interesting physics (see e.g. \cite{Bianchi:2014qua, Bianchi:2014vea,Good:2017ddq,Cong:2018vqx,Romualdo:2019eur,Lee:2019adw,Good:2019tnf,Foo_2020}). Since the stress tensor is technically divergent in the model, a point-splitting regularization technique is used to determine finite meaningful particle production results (see e.g. \cite{Stargen:2016euf,Good:2020uff}). 

Based on this 50 year old prescription, the particle production and energy flux are generally thought to be a consequence of the mirror’s acceleration and jerk, respectively \cite{Birrell1982,Fabbri}. Moreover, the close relation with thermodynamics has been analyzed (see e.g. \cite{Davies:1982cn,Helfer:2000fg}), as a product of the entanglement, entropy, and relativistic quantum information issues associated with the effect (see e.g. \cite{Giulio,Chen:2017lum,Good:2018aer}). Physical detection is an ongoing priority \cite{Chen:2015bcg,Chen:2020sir}, especially since superconducting quantum interference devices act as moving mirrors and Bose-Einstein condensates have demonstrated experimental success for investigation of the dynamical Casimir effect \cite{Dodonov:2020eto}.

The main approach in this paper is the use of an accelerated boundary correspondence in flat spacetime, which creates particles in a fundamentally related way to the radiation in a curved spacetime. In flat spacetime the particle production is caused by the dynamical acceleration of the boundary (Davies-Fulling effect \cite{Davies:1976hi,Davies:1977yv}).\footnote{In flat spacetime there also exists particle creation due to acceleration without the presence of a boundary: an accelerating observer detects particles in flat spacetime (Unruh effect \cite{Unruh:1976db}).} In curved spacetime the particle production is caused by dynamical gravitational fields in the formation of black holes (Hawking effect \cite{Hawking:1974sw}), but it can also be caused by the acceleration of observers to remain static around black holes (Unruh effect \cite{Unruh:1976db}) or in an expanding universe (Gibbons-Hawking effect \cite{GibbonsHawking}). Accordingly, the moving mirror analogy is a simple and useful way to study and understand the quantum effects produced by gravity. This is the aim of solving for the accelerated boundary correspondences (ABCs). The associated mirrors have perfect reflection, and although these models are limited in scope (e.g. no backscattering, grey-body effect, superradiance, etc.), the solutions successfully model the `pure' quantum particle creation effects in either black hole spacetimes or even cosmological spacetimes. This implies that only the s-waves of `pure' gravitational particle production are accounted by this kind of mirrors \cite{Fabbri}.

In this context, the ABCs for the Schwarzschild black hole \cite{Good:2016oey}, the Reissner-Nordström black hole \cite{Good:2020qsy}, the Kerr black hole \cite{Good:2020fjz} (and the Kerr-Newman black hole \cite{Foo:2020bmv}), and the Taub-NUT black hole \cite{Foo:2020xmy} have been recently solved. The range of ABC utility is quite broad, as particular asymptotic dynamics characterize well-known non-thermal curved spacetime black hole end-states, including:
\begin{itemize}
    \item extremals (uniform acceleration \cite{Liberati:2000sq,Rothman:2000mm,good2020extreme,Good:2020fjz,Foo:2020bmv}),
    \item remnants (constant-velocity  \cite{Good:2015nja,Good:2016atu,Good:2016yht,Good:2018zmx,Myrzakul:2018bhy,Good:2018ell}),
    \item effervescents (zero-velocity  \cite{Walker_1982,Good:2017ddq,Good:2018aer,Good:2019tnf,B,Good:2017kjr,GoodMPLA}).
\end{itemize}
Even the de Sitter and anti-de Sitter cosmology have a respective ABC \cite{Good:2020byh}. Despite this progress, the amplification of quantum vacuum fluctuations that result in the all-time particle spectrum (static observers) for the formation of a black hole embedded in an expanding universe have not been resolved, in the sense that the Bogoliubov coefficients are unknown. That is to say, the SdS moving mirror has also not been explored, even though the Schwarzschild mirror \cite{Good:2016oey} and de Sitter mirror \cite{Good:2020byh} Bogoliubov coefficient solutions are known. In this paper we analyze the particle creation by the SdS mirror directly, obtaining analytical expressions for the Bogoliubov coefficients. The analog SdS mirror will describe the results detected by static observers near the (future) cosmological horizon of SdS spacetime. We combine our understanding of both the Schwarzschild black hole and the de Sitter cosmology to derive a particle spectrum which is consistent in both asymptotic regimes. This contribution is a demonstration and touchstone of the established program where the simplicity of analog gravitational systems in a flat spacetime with lower dimensions have shed light on the more complicated curved spacetime higher dimensional physics.  This approach is expected to be tractable in part because in (1+1)-dimensional SdS spacetime one can analytically solve the wave equation \cite{Markovic:1991ua, Anderson:2020dim}, where there is no effective potential and so the effects of the causal structure on the initial waveforms are accentuated. Despite significant simplifications that take place in (1+1) dimensions, nontrivial quantum effects are present.

The results in the above mentioned papers suggest that the properties of the radiation of SdS spacetime combination are qualitatively different from those of the two constituents. This is one motivation to concentrate on the radiation solution describing the asymptotically de Sitter Schwarzschild black hole with respect to static observers. In agreement with \cite{cd}, which used energy conservation and self-gravitation interaction, our derived all-time SdS particle spectrum determined is explicitly shown to be non-thermal (see also \cite{Choudhury:2004ph}), although this work confirms two asymptotic Planck distributions (at their respective temperatures).

The paper is organized as follows: in Sec.~\ref{Sec2} we briefly introduce the SdS spacetime, obtaining the horizons as well as the corresponding surface gravities. We also solve for the Regge-Wheeler (tortoise) coordinate. In Sec.~\ref{Sec3} we deal with the SdS moving mirror by correctly interpreting and mapping the SdS spacetime into the corresponding flat spacetime plus moving mirror equation of motion. In Sec.~\ref{Sec4} we demonstrate that the energy flux of this system is consistent with the known physics of the SdS spacetime black hole and cosmological horizons. In Sec.~\ref{Sec5} the particle production results are shown. In Sec.~\ref{Sec6} we briefly investigate the corresponding spacetime associated with the approximated asymptotic SdS moving mirror. In Sec.~\ref{Sec7} we conclude. Throughout we use natural units: $G=c=\hbar=k_B =1$.

\section{Schwarzschild-de Sitter Spacetime}\label{Sec2}
We consider the form of a general spherically symmetric static spacetime. In spherical coordinates, the metric reads
\be
    ds^2=-f(r)dt^2+f(r)^{-1}dr^2+r^2d\Omega^2,\label{SS}
\ee
with $d\Omega^2=d\theta^2+\sin^2\theta d\varphi^2$. The SdS metric is the simplest solution describing a black hole embedded in an expanding universe with positive cosmological constant\footnote{The SdS metric is otherwise known as the Kottler metric \cite{Kottler1918}.}. It follows from Eq.~(\ref{SS}) with
\be
    f(r)\equiv f_{SdS}=1-\frac{2M}{r}-\frac{\Lambda r^2}{3},
\ee
where $M$ is the black hole mass and $\Lambda>0$ is the cosmological constant \cite{Stephani2003}. One can directly verify that the Schwarzschild case is recovered for $\Lambda=0$ and the de Sitter case is recovered for $M=0$. All along, we will use the parameter $L\equiv\sqrt{3/\Lambda}$, so that
\be
    f(r)\equiv f_{SdS}=1-\frac{2M}{r}-\frac{r^2}{L^2}.\label{f_SdS1}
\ee
By setting $f_{SdS}=0$, and also assuming that $3\sqrt{3}M/L<1$, the black hole and cosmological horizons are, respectively, at $r=r_B$ and $r=r_C$ with  
\be
    r_B \equiv \frac{2L}{\sqrt{3}}\cos\left[\frac{1}{3}\cos^{-1}\left(\frac{3\sqrt{3}M}{L}\right)+\frac{\pi}{3}\right],
\ee
and
\be
    r_C \equiv \frac{2L}{\sqrt{3}}\cos\left[\frac{1}{3}\cos^{-1}\left(\frac{3\sqrt{3}M}{L}\right)-\frac{\pi}{3}\right],
\ee
where $r_B<r_C$. There also exists a negative, nonphysical solution given by $r=r_0\equiv-(r_B+r_C)$. This solution is unphysical since one cannot extend the coordinate range beyond the curvature singularity at $r=0$. In terms of these parameters we rewrite Eq.~(\ref{f_SdS1}) as
\be
    f_{SdS}=\frac{(r-r_B)(r_C-r)(r-r_0)}{L^2r},\label{f_SdS2}
\ee
which imply the following relations:
\be
    \begin{split}
        M&=\frac{(r_B+r_C)r_Br_C}{2(r_B^2+r_C^2+r_Br_C)},\\
        L^2&=r_B^2+r_C^2+r_Br_C.
    \end{split}
\ee
The surface gravity associated with each horizon $r_i$ is defined from the relation $\kappa_i=\frac{1}{2}\left|df_{SdS}/dr\right|_{r=r_i}$. Explicitly, we get
\be
    \begin{split}
        \kappa_B&=\frac{(r_C-r_B)(r_B-r_0)}{2L^2r_B},\\
        \kappa_C&=\frac{(r_C-r_B)(r_C-r_0)}{2L^2r_C},\\
        \kappa_0&=-\frac{\kappa_B\kappa_C}{\kappa_C-\kappa_B}.
	\end{split}
\ee
Finally, we will be interested in writing the SdS metric in a double null (light-cone) coordinate system $(u,v)$ and, for that, we will need the Regge-Wheeler or tortoise coordinate $r^*$ defined by $r^*\equiv\int f_{SdS}^{-1}\,dr$. The horizons $r_i$ and surface gravities $\kappa_i$ allow us to write $f_{SdS}^{-1}$ from Eq.~(\ref{f_SdS2}) simply as
\be
    f_{SdS}^{-1}=\frac{1}{2\kappa_B(r-r_B)}+\frac{1}{2\kappa_C(r_C-r)}+\frac{1}{2\kappa_0(r-r_0)},
\ee
thence
\be
    \begin{split}
        r^*&=\frac{1}{2\kappa_B}\ln\left(\frac{r-r_B}{r_B}\right)-\frac{1}{2\kappa_C}\ln\left(\frac{r_C-r}{r_C}\right)\\
        &+\frac{1}{2\kappa_0}\ln\left(-\frac{r-r_0}{r_0}\right).\label{r*}
    \end{split}
\ee
The Regge-Wheeler or tortoise coordinate will prove to be more than convenience; it is physically related with the trajectory of the moving mirror in the analog gravitational system.

\section{Trajectory Dynamics}\label{Sec3}
We are now ready to solve the accelerated boundary correspondence with SdS spacetime. This is achieved by considering the matching of two metrics. This matching condition appears when one considers the dynamical gravitational model of a null shell collapsing in a (1+3)-dimensional spacetime. A (1+3)-dimensional spacetime of this type allows us to deduce the corresponding trajectory for the perfectly reflecting SdS mirror analog in (1+1)-dimensional flat spacetime. 

\subsection{Geometric considerations of the (1+3)-dimensional gravitational collapse of a null shell}
The (1+3)-dimensional spacetime describing the dynamical gravitational collapse of a null shell at $v=v_0$ takes advantage of matching two metrics: the metric inside the shell ($v<v_0$) and the metric outside the shell ($v>v_0$) \cite{Fabbri}. In this case, we apply the matching along a null shell between an inside (``in'') Minkowski metric and an outside (``out'') SdS metric.

The ``in'' metric is
\be
    ds^2=-dt_{\mathrm{in}}^2+dr^2+r^2d\Omega^2,
\ee
which we rewrite in the double null coordinate system $(u_{\mathrm{in}},v)$ with $u_{\mathrm{in}}=t_{\mathrm{in}}-r$ and $v=t_{\mathrm{in}}+r$ as
\be
    ds^2=-du_{\mathrm{in}}dv+r^2d\Omega^2.
\ee
The ``out'' metric is
\be
    ds^2=-f_{SdS}dt_{\mathrm{out}}^2+f_{SdS}^{-1}dr^2+r^2d\Omega^2,
\ee
with $f_{SdS}$ given by Eq.~(\ref{f_SdS1}) or Eq.~(\ref{f_SdS2}), which we rewrite in the double null coordinate system $(u_{\mathrm{out}},v)$ with $u_{\mathrm{out}}=t_{\mathrm{out}}-r^*$ and $v=t_{\mathrm{out}}+r^*$ with $r^*$ given by Eq.~(\ref{r*}) as
\be
    ds^2=-f_{SdS}du_{\mathrm{out}}dv+r^2d\Omega^2.
\ee
The matching condition between both ``in'' and ``out'' metrics\footnote{Note that the ``out'' metric is the SdS metric written in static coordinates. As a consequence, the analog SdS mirror trajectory in Minkowski spacetime will represent static observers in SdS spacetime.} demands the continuity of the coordinates $r$ and $v$ along the null-shell $(v=v_0)$\footnote{This is why we add no subscripts for the coordinates $r$ and $v$.} and reads simply as
\be
    r(u_{\mathrm{in}}, v_0)=r(u_{\mathrm{out}},v_0).
\ee
Accordingly, if we write
\be
    r(u_{\mathrm{in}},v_0)=\frac{1}{2}(v_0-u_{\mathrm{in}})
\ee
and
\be
    r^*(u_{\mathrm{out}},v_0)=\frac{1}{2}(v_0-u_{\mathrm{out}}),
\ee
we arrive at the coordinate relation $u_{\mathrm{out}}=u_{\mathrm{out}}(u_{\mathrm{in}})$ through Eq.~(\ref{r*}),
\be
    \begin{split}
        u_{\mathrm{out}}(u_{\mathrm{in}})=v_0 &-\frac{1}{\kappa_B}\ln\left(\frac{v_0-u_{\mathrm{in}}-2r_B}{2r_B}\right)\\
        &+\frac{1}{\kappa_C}\ln\left(\frac{2r_C-v_0+u_{\mathrm{in}}}{2r_C}\right)\\
        &-\frac{1}{\kappa_0}\ln\left(-\frac{v_0-u_{\mathrm{in}}-2r_0}{2r_0}\right).
    \end{split}
\ee
From the above expression we check that if $u_{\mathrm{in}}=v_0-2r_B$ we have that $u_{\mathrm{out}}\rightarrow +\infty$ whereas if $u_{\mathrm{in}}=v_0-2r_C$ we have that $u_{\mathrm{out}}\rightarrow-\infty$. Hence, we define $v_B\equiv v_0-2r_B$ and $v_C\equiv v_0-2r_C$, denoting the null rays that represent the horizons of SdS spacetime. The coordinate relation $u_{\mathrm{out}}=u_{\mathrm{out}}(u_{\mathrm{in}})$ is, then,
\be
    \begin{split}
        u_{\mathrm{out}}(u_{\mathrm{in}})=v_0 &-\frac{1}{\kappa_B}\ln\left(\frac{v_B-u_{\mathrm{in}}}{2r_B}\right)\\
        &+\frac{1}{\kappa_C}\ln\left(\frac{u_{\mathrm{in}}-v_C}{2r_C}\right)\\
        &-\frac{1}{\kappa_0}\ln\left(-\frac{v_0-u_{\mathrm{in}}-2r_0}{2r_0}\right),
    \end{split}
\ee
with $v_C<u_{\mathrm{in}}<v_B$. We can set (without loss of generality) $v_0=0$ so $v_B=-2r_B$, $v_C=-2r_C$, and
\be
    \begin{split}
        u_{\mathrm{out}}(u_{\mathrm{in}})= &-\frac{1}{\kappa_B}\ln\left(\frac{v_B-u_{\mathrm{in}}}{2r_B}\right)\\
        &+\frac{1}{\kappa_C}\ln\left(\frac{u_{\mathrm{in}}-v_C}{2r_C}\right)\\
        &-\frac{1}{\kappa_0}\ln\left(\frac{u_{\mathrm{in}}+2r_0}{2r_0}\right).
    \end{split}
\ee

\subsection{(1+1)-dimensional SdS mirror trajectory}
\begin{figure}
    \centering
    \includegraphics[width=0.9\linewidth]{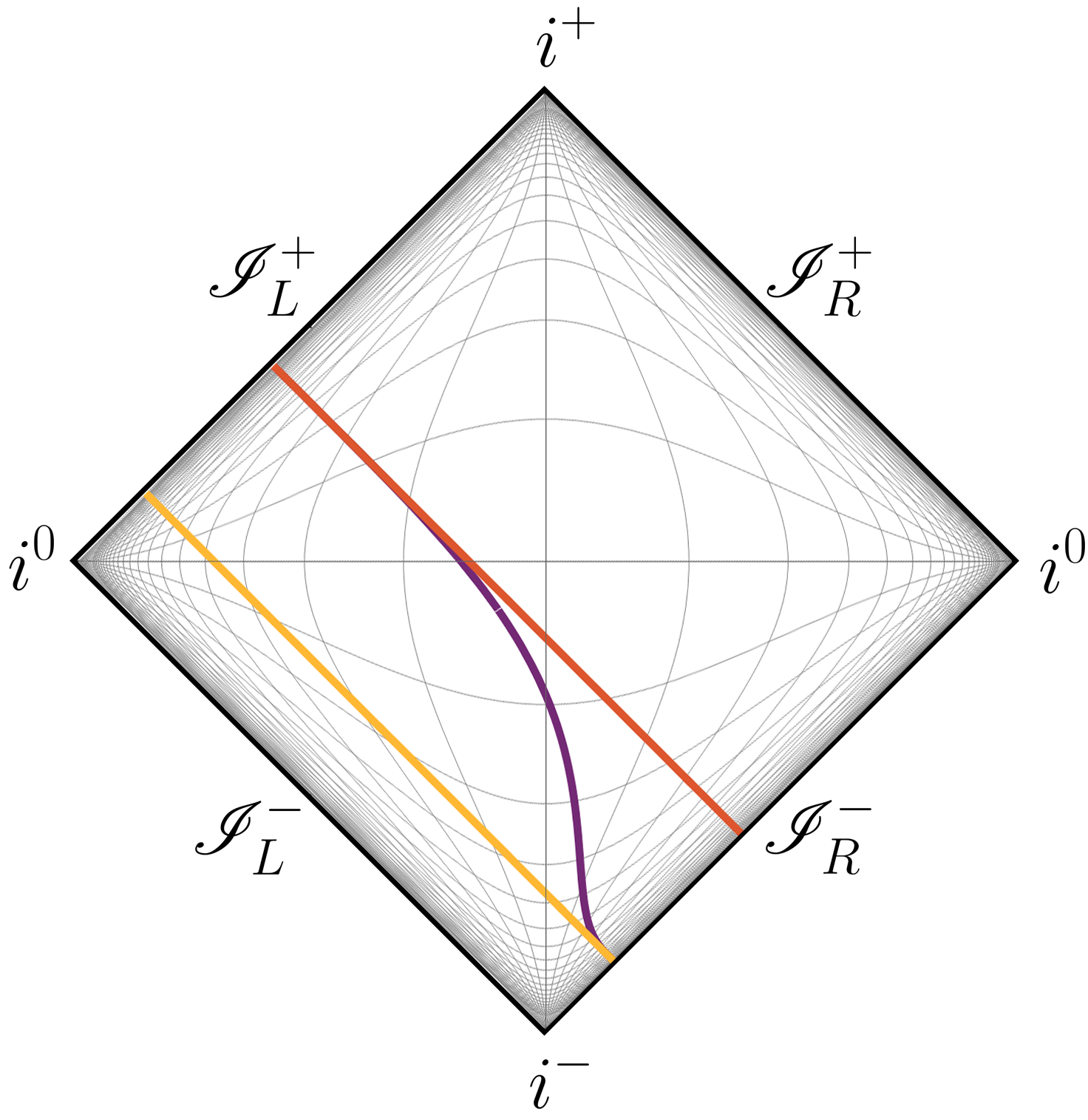}
    \caption{Penrose diagram showing the trajectory, Eq.~(\ref{SdSm2}), of the SdS moving mirror. We see the two acceleration horizons (yellow and red lines), which are analogs of the past cosmological horizon and the future black hole horizon, respectively. The purple line is the SdS moving mirror. Here the gravitational parameters are taken as $M=1/24$ and $L=2/3$ for clarity.}
    \label{Fig1}
\end{figure}

A massless quantum scalar field $\phi(x)$ propagating in a (1+3)-dimensional spacetime with spherical symmetry must satisfy the condition $\phi(t,r=0)=0$, since the correspondent modes are spherical-type waves. This regularity condition at $r=0$ allows determination of the exact form of the ``out'' modes in the ``in'' (Minkowski) part of the spacetime where at $r=0$ we identify the (Minkowski) coordinates $u_{\mathrm{in}}$ and $v$, i.e.  $u_{\mathrm{in}}=v$. On the other hand, a massless quantum scalar field $\phi(x)$ propagating in (1+1)-dimensional Minkowski spacetime with a mirror must satisfy the boundary condition $\phi(t,z(t))=0$, where $z(t)$ parameterizes the mirror perfectly reflecting trajectory. Accordingly, we can think of the origin of coordinates $r=0$ in the (1+3)-dimensional spacetime as a mirror in flat spacetime with a trajectory that we can read straightforwardly as the the coordinate relation $u_{\mathrm{out}}=u_{\mathrm{out}}(v)$. At this point, we move to the (1+1)-dimensional acceleration model. The SdS moving mirror is completely described by the following function of advanced time that parameterizes the perfectly reflecting trajectory in flat spacetime:
\be
    \begin{split}
       f(v)=&-\frac{1}{\kappa_B}\ln\left(\frac{v_B-v}{2r_B}\right)
       \\
       &+\frac{1}{\kappa_C}\ln\left(\frac{v-v_C}{2r_C}\right)\\
       &-\frac{1}{\kappa_0}\ln\left(\frac{v+2r_0}{2r_0}\right),\label{SdSm1}
    \end{split}
\ee
Finally, let us rewrite the SdS mirror trajectory, Eq.~(\ref{SdSm1}), as
\be
    \begin{split}
       f(v)=&-\frac{1}{\kappa_B}\ln\left[\Bar{\kappa}_B(v_B-v)\right]
       \\
       &+\frac{1}{\kappa_C}\ln\left[\Bar{\kappa}_C(v-v_C)\right]\\
       &-\frac{1}{\kappa_0}\ln\left[\Bar{\kappa}_0(v+v_B+v_C)\right],\label{SdSm2}
    \end{split}
\ee
where the constants $\Bar{\kappa}_B\equiv(2r_B)^{-1}$, $\Bar{\kappa}_C\equiv(2r_C)^{-1}$ and $\Bar{\kappa}_0\equiv(2r_0)^{-1}$ have been conveniently defined. In the SdS mirror trajectory, Eq.~(\ref{SdSm2}), one should understand $\kappa_B$ (or $\kappa_C$) as an acceleration parameter in the future (or in the past), and $v_B$ (or $v_C$) as the null asymptote of the mirror trajectory in the future (or in the past). In gravitational terms, $\kappa_B$ (or $\kappa_C$) is the black hole (or cosmological) surface gravity and $v_B$ (or $v_C$) is the null ray representing the black hole (or cosmological) horizon. The causal history of Eq.~(\ref{SdSm2}) is depicted in a Penrose diagram in Fig.~\ref{Fig1}.

As a check, one can evaluate the Schwarzschild limit $(L\rightarrow\infty)$ and de Sitter limit $(M\rightarrow0)$ of the accelerated boundary correspondence with SdS spacetime, Eq.~(\ref{SdSm2}). In the limit $L\rightarrow\infty$ we get
\be
    f(v)\simeq v-\frac{1}{\kappa}\ln\left[\kappa(v_H-v)\right],
\ee
with $\kappa=(4M)^{-1}$ and $v_H=-4M$, recovering the result of the accelerated boundary correspondence with the Schwarzschild black hole \cite{Good:2016oey}. In the limit $M\rightarrow0$ we get
\be
    f(v)\simeq\frac{2}{\kappa}\tanh^{-1}\left(\frac{\kappa v}{2}\right),
\ee
where $\kappa=L^{-1}$ and $v_H=\pm2L$, recovering the result of the accelerated boundary correspondence with de Sitter cosmology \cite{Good:2020byh}.

\section{Energy Flux}\label{Sec4}
\begin{figure}
    \centering
    \includegraphics[width=\linewidth]{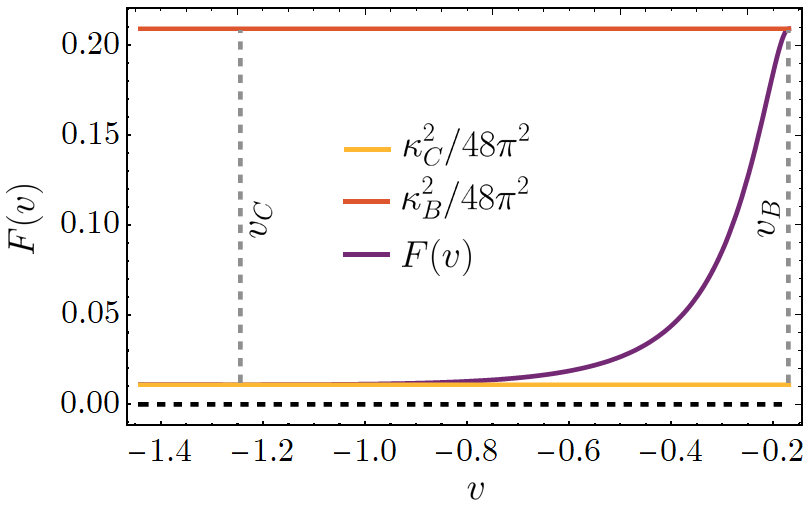}
    \caption{Energy flux, Eq.~(\ref{EF}), as a function of advanced time $v$; demonstrating the different constant energy flux at the horizon locations (yellow and red lines). Here the gravitational parameters are taken as $M=1/24$ and $L=2/3$ for consistency with Fig. \ref{Fig1}. The key takeaway is that the energy flux (purple line) increases monotonically from the smaller constant value of the analog cosmological horizon in the past to the larger constant value of the analog black hole horizon in the future. The black hole horizon is hotter than the cosmological horizon. See Eqs.~(\ref{Fc}) and (\ref{Fb}) for the early-time and late-time energy flux asymptotic behavior.} 
    \label{Fig2}
\end{figure}

After having analyzed the accelerated boundary correspondence with SdS spacetime and before studying the particle creation by the SdS moving mirror, we find it insightful to investigate the energy flux radiated by the perfectly reflecting trajectory. An analysis of the energy-momentum tensor in the (1+1)-dimensional quantum theory of a massless scalar field disturbed by the boundary dynamics in flat spacetime yields an expression for the energy flux \cite{Davies:1976hi, Davies:1977yv}, which one can write as
\be
    F(v)=\frac{1}{24\pi}\left(\frac{df(v)}{dv}\right)^{-2}\left\lbrace f(v),v\right\rbrace\label{EF}.
\ee
The brackets define the Schwarzian derivative,
\be
    \left\lbrace f(v),v\right\rbrace\equiv\frac{f'''(v)}{f'(v)}-\frac{3}{2}\left(\frac{f''(v)}{f'(v)}\right)^2,
\ee
where the prime denotes derivative with respect to $v$. The explicit expression that follows from Eqs.~(\ref{SdSm2}) and (\ref{EF}) is lengthy but, from it, we can get an idea of the constant energy flux and the distribution of particles near the acceleration horizons (null asymptotes) of the mirror trajectory. A plot of the energy flux is given in Fig. \ref{Fig2}. On the one hand, at early times $(v\rightarrow v_C)$ and to leading order we have that
\be
    F(v)=\frac{\kappa_C^2}{48\pi}+\mathcal{O}((v-v_C)^2).\label{Fc}
\ee
On the other hand, at late times $(v\rightarrow v_B)$ and to leading order we have that
\be
    F(v)=\frac{\kappa_B^2}{48\pi}+\mathcal{O}((v-v_B)^2).\label{Fb}
\ee
These results show different constant energy flux at early and late times for the SdS moving mirror. Furthermore, the above results, Eqs.~(\ref{Fc}) and (\ref{Fb}) suggest two Planck distributions of particles at different temperatures $T_C$ and $T_B$ that are proportional to the acceleration parameter in the past and future $\kappa_C=2\pi T_C$ and $\kappa_B=2\pi T_B$, respectively. These two thermal spectra will be explicitly calculated in the next section.

\section{Particle Production}\label{Sec5}
\begin{figure}
    \centering
    \includegraphics[width=0.9\linewidth]{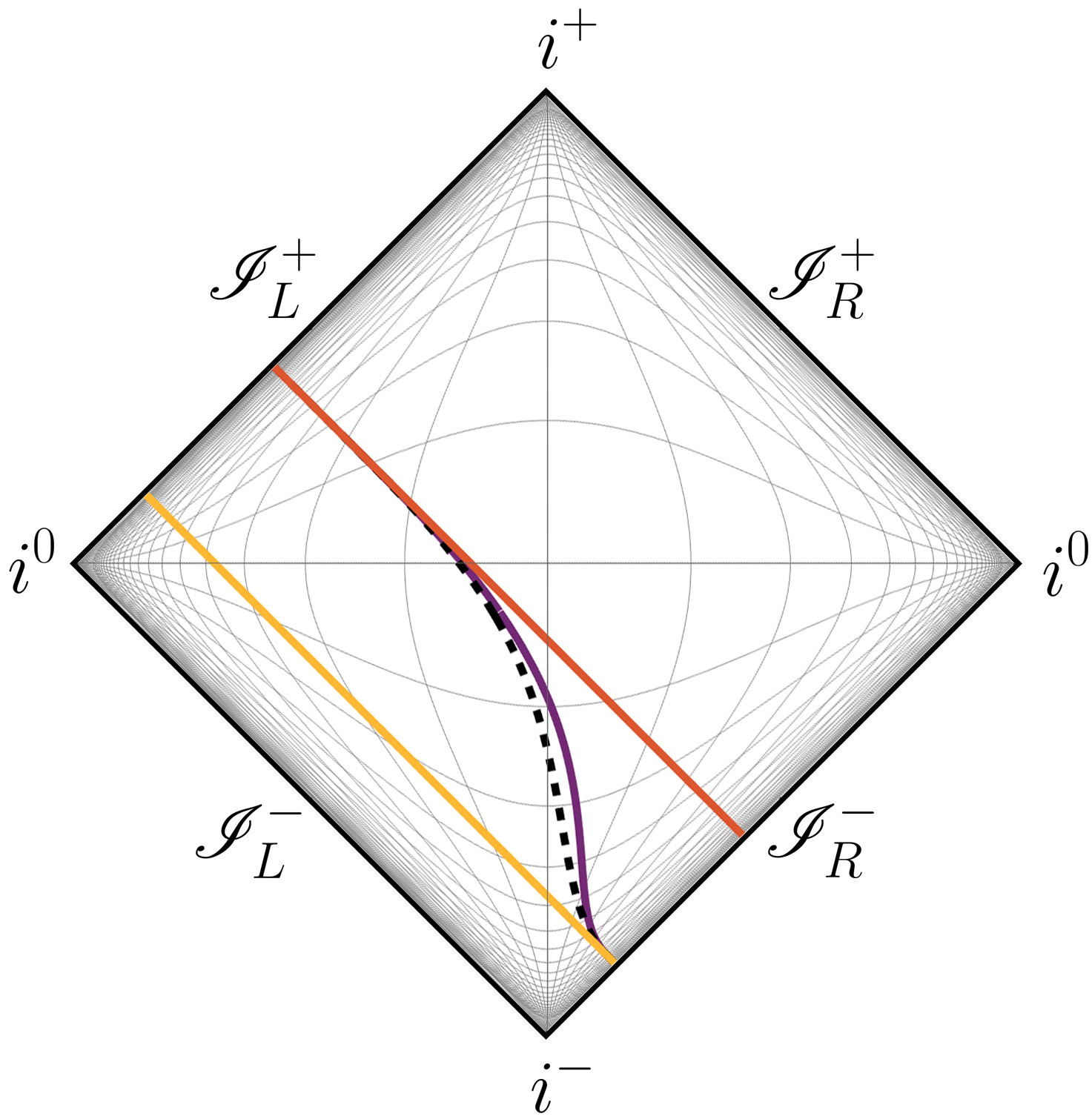}
    \caption{Penrose diagram showing the approximated and exact SdS moving mirror trajectories, Eq.~(\ref{SdSma}) (black dashed line) and Eq.~(\ref{SdSm2}) (purple line).  Wee see that asymptotic version of the SdS mirror trajectory also has the exact same two horizons, which are analogs of the past cosmological horizon (yellow line) and the future black hole horizon (red line). Here $M=1/24$ and $L=2/3$ as before. At the intermediate times of acceleration the bifurcation of both trajectories manifests.}
    \label{Fig3}
\end{figure}

\begin{figure}
    \centering
    \includegraphics[width=\linewidth]{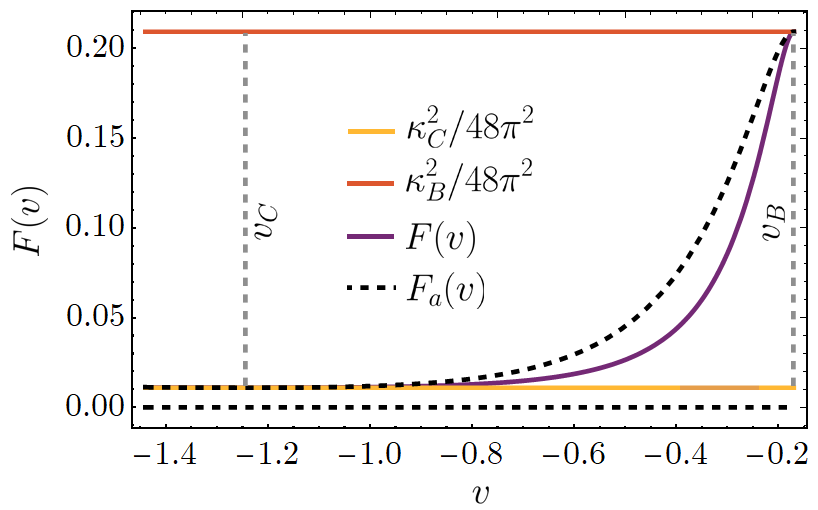}
    \caption{Energy fluxes $F_a(v)$ and $F(v)$ for the approximated and exact SdS moving mirror trajectories, Eqs.~(\ref{SdSma}) and (\ref{SdSm2}), obtained from Eq.~(\ref{EF}) as a function of advanced time $v$; demonstrating the different constant energy fluxes at the horizon locations (yellow and red lines). Here $M=1/24$ and $L=2/3$ as before. The key takeaway is that the energy fluxes, $F_a(v)$ and $F(v)$, are similar. Again, like in Fig. \ref{Fig2}, the energy flux increases monotonically from the smaller constant value of the analog cosmological horizon in the past to the larger constant value of the analog black hole horizon in the future, highlighting that the black hole horizon is hotter than the cosmological horizon. Eqs.~(\ref{Fc}) and (\ref{Fb}) are the usual early-time and late-time energy flux asymptotic behavior. The same color scheme is used in the Penrose diagram of Fig. \ref{Fig3}.}
    \label{Fig4}
\end{figure}

We are now in a position to address the particle creation by the SdS moving mirror. The main goal is to obtain the solution to the Bogoliubov coefficient $\beta_{\omega\omega'}$ \cite{Birrell1982},
\be
    \beta_{\omega\omega'}=\frac{1}{2\pi}\sqrt{\frac{\omega'}{\omega}}\int_{v_C}^{v_B}dv\,e^{-i\omega'v-i\omega f(v)},\label{beta}
\ee
where $\omega$ $(\omega')$ are the frequencies of the outgoing (ingoing) modes. The calculation of the beta Bogoliubov coefficient directly via Eq.~(\ref{SdSm2}) is intractable. As a first approximation, let us consider the following form for the SdS mirror trajectory:
\be
    f(v)\approx-\frac{1}{\kappa_B}\ln\left[\Bar{\kappa}_B(v_B-v)\right]+\frac{1}{\kappa_C}\ln\left[\Bar{\kappa}_C(v-v_C)\right],\label{SdSma}
\ee
instead of the full expression, Eq.~(\ref{SdSm2}). Note that we have ignored the third logarithmic term. This approximation of the SdS mirror trajectory, Eq.~(\ref{SdSma}), finds its justification at early and late times, as it presents the same asymptotic behaviour in the far past and far future as the exact expression. Notice that, in the exact SdS mirror trajectory, Eq.~(\ref{SdSm2}), the first logarithmic term dominates at late times $(v\rightarrow v_B)$ and the second logarithmic term dominates at early times $(v\rightarrow v_C)$, which are the two terms in the approximated SdS mirror trajectory, Eq.~(\ref{SdSma}). The third logarithmic term has significance not at early and late times but at the intermediate times of the trajectory. However, in the regime $3\sqrt{3}M/L\simeq1$ (which we have considered in Figs.\ \ref{Fig3} and \ref{Fig4}), one finds that the magnitude of this third term is negligible in comparison to the first and second terms, justifying our approximation. See Fig. \ref{Fig3} for a Penrose diagram with the approximated trajectory overlaid on the exact trajectory. See Fig. \ref{Fig4} for a comparison of the energy fluxes of the approximated trajectory overlaid on the energy flux of the exact trajectory. As shown in Figs.\ \ref{Fig3} and \ref{Fig4}, the asymptotic and exact SdS mirrors are qualitatively similar (both dynamically and energetically), and hence one can have confidence that they will likewise possess a similar particle production spectra as a result.

This approximation is inspired by the method used in the first papers on analog moving mirrors by Fulling and Davies \cite{Davies:1976hi, Davies:1977yv}. In them, they began by considering an asymptotic form of a mirror trajectory at late times since the behaviour of the mirror trajectory at early times was considered irrelevant to analyze the particle production in the asymptotic late-time region. What Fulling and Davies did was to approximate the calculation of the Bogoliubov coefficients by using the asymptotic late-time form of the mirror trajectory at all times. Then, the beta Bogoliubov coefficient derived in that way was a good approximation in describing the particle production in the asymptotic late-time region. That analog mirror asymptotic solution has the same form as the asymptotic Schwarzschild mirror. In the case of the analog SdS moving mirror developed in this paper, we have determined the exact form of the mirror trajectory, Eq.~(\ref{SdSm2}), but since the computation of the Bogoliubov coefficients is intractable, and the asymptotic early-time and late-time regions are relevant to study the particle production, we approximate the mirror trajectory by using the asymptotic form, Eq.~(\ref{SdSma}), at all times; just as Fulling and Davies did. Now, the beta Bogoliubov coefficient to be derived in this way will be a good approximation in describing the particle production in the asymptotic early-time and late-time regions. Importantly, the asymptotic SdS mirror will certainly give approximate but physically relevant particle production results in the intermediate-time region.

Thus, bearing in mind the regimes of validity of the approximation and the subsequent discussion, we will derive the beta Bogoliubov coefficient from Eq.~(\ref{beta}) with $f(v)$ given by Eq.~(\ref{SdSma}). Then, we define
\be
    a\equiv\frac{i\omega'}{\kappa'}\quad,\quad b\equiv\frac{i\omega}{\kappa_B},\quad,\quad c\equiv\frac{i\omega}{\kappa_C}\quad.\label{abc}
\ee
with $\kappa'=(v_B-v_C)^{-1}$. The final result is
\be
    \begin{split}
      \beta_{\omega\omega'}=&\frac{1}{2\pi}\sqrt{\frac{\omega'}{\omega}}\frac{(\Bar{\kappa}_B)^b(\Bar{\kappa}_C)^{-c}}{(\kappa')^{1+b-c}}e^{-i\omega'v_B}\\
      &\cdot\frac{\Gamma(1+b)\Gamma(1-c)}{\Gamma(2+b-c)}{}_1F_1(1+b;2+b-c;a),\label{betaSdSma1}
    \end{split}
\ee
where $\Gamma(x)$ is the Gamma function and ${}_1F_1(p;q;z)$ is the confluent hypergeometric function of the first kind. The explicit derivation of the above expression is developed in Appendix \ref{A}.  Thereby, using the regularized confluent hypergeometric function ${}_1\widetilde{F}_1(p;q;z)\equiv{}_1F_1(p;q;z)/\Gamma(q)$, we obtain
\be
    \begin{split}
      \beta_{\omega\omega'}=&\frac{1}{2\pi}\sqrt{\frac{\omega'}{\omega}}\frac{(\Bar{\kappa}_B)^b(\Bar{\kappa}_C)^{-c}}{(\kappa')^{1+b-c}}e^{-i\omega'v_B}\\
      &\cdot\Gamma(1+b)\Gamma(1-c){}_1\widetilde{F}_1(1+b;2+b-c;a).\label{betaSdSma2}
    \end{split}
\ee
The particle spectrum per mode $\omega$ and per mode $\omega'$ is the modulus square of the beta Bogoliubov coefficient in Eq.~(\ref{betaSdSma2}), 
\be
    N_{\omega\omega'}\equiv\left|\beta_{\omega\omega'}\right|^2.
\ee
Therefore, we write $N_{\omega\omega'}$ as
\be
N_{\omega\omega'}= \frac{a^2 b c}{\omega\omega'}\frac{e^{i \pi  (b+c)}}{\left(e^{2 i \pi  b}-1\right) \left(e^{2 i \pi  c}-1\right)}\left|_1\widetilde{F}_1\right|^2,\label{N}
\ee
Notice the not-so-inconspicuous hint of two Planck distributions in the denominator. In the high frequency regime, i.e. $\omega'\gg\omega$, we expect to see a late-time thermal spectrum, as is the case in Hawking radiation \cite{Hawking:1974sw}, see e.g. the analog Schwarzschild black hole mirror \cite{Good:2016oey}. Importantly, in this situation we also expect to see a early-time thermal spectrum, corresponding the de Sitter cosmological horizon. Expanding Eq.~(\ref{N}) for large $\omega'$, gives to leading order,
\be
N_{\omega\omega'} = N_{B} + N_{C} + N_{BC},
\ee
where 
\be
N_{B} \equiv \frac{1}{2\pi \kappa_B \omega'}\frac{1}{e^{2\pi \omega/\kappa_B}-1},\label{NB}
\ee
which is the expected late-time thermal spectrum for the black hole horizon. This result agrees with particle creation by the asymptotic version of the SdS moving mirror in a Planck distribution at a temperature $T_B$ that is proportional to acceleration parameter of the boundary trajectory in the future, $\kappa_B=2\pi T_B$. What about the expected early-time thermal spectrum for the cosmological horizon? This term also makes an appearance:
\be
N_{C} \equiv \frac{1}{2\pi \kappa_C \omega'}\frac{1}{e^{2\pi \omega/\kappa_C}-1}.\label{NC}
\ee
This result agrees with particle creation by the asymptotic version of the SdS moving mirror in a Planck distribution at a temperature $T_C$ that is proportional to acceleration parameter of the boundary trajectory in the past, $\kappa_C=2\pi T_C$. With the above arguments, we confirm that the (Schwarzschild black hole) late-time thermal spectrum in Eq.~(\ref{NB}) and the (de Sitter cosmology) early-time thermal spectrum in Eq.~(\ref{NC}) are present in SdS moving mirror system. Furthermore, an additional complicated cross term is obtained, indicating the approximate form of the interaction between both the black hole horizon and the cosmological horizon of SdS spacetime. The presence of this term is reminiscent of the treatment of radiation resistances for resistors in `parallel' (see e.g. \cite{Gray:2015pma}). This term is
\be
    N_{BC}=\Bar{N}_{BC}+\Bar{N}_{BC}^*,
\ee
where
\be
    \begin{split}
        \Bar{N}_{BC}\equiv&-\frac{1}{4\pi^2\omega\omega'} \left(\frac{\omega'}{\kappa'}\right)^{-\frac{i\omega}{\kappa}}e^{-\frac{\pi\omega}{2\kappa}-\frac{i\omega'}{\kappa'}}\\
        &\cdot\Gamma\left(1+\frac{i\omega}{\kappa_B}\right)\Gamma\left(1+\frac{i\omega}{\kappa_C}\right),
    \end{split} \label{inter}
\ee
where we have defined $\kappa^{-1}=\kappa_B^{-1}+\kappa_C^{-1}$, and $\kappa' = (v_B - v_C)^{-1}$ as in Eq.~(\ref{abc}). This result indicates the particle creation by the asymptotic version of the SdS moving mirror at intermediate times. It demonstrates a non-thermal particle distribution between the early-time and late-time thermal equilibria. Note that this result implies (for SdS) that the regime $\omega'\gg \omega$ captures both the late time and the early time radiation (this limit is most commonly associated with late-time radiation in the context of Schwarzschild \cite{Hawking:1974sw}). Eq.~(\ref{inter}) has a counterpart in the SdS moving mirror system, and it should take a different but similar form since it shows the effect of the acceleration horizons and the interaction behaviour between them resulting in the non-thermal spectrum. The radiation will almost certainly be influenced by the intermediate-time trajectory dynamics. Eq.~(\ref{inter}) is one of the more interesting results of our analysis and a key novelty of the SdS-type radiation where two Planck distributions `interact' to give an overall non-thermal spectrum. 

\section{Dual-Temperature ABC}\label{Sec6}
Once having obtained the early-time and late-time particle production results for the SdS moving mirror through the use of the trajectory Eq.~(\ref{SdSma}), an open question remains: What is the approximate form of SdS spacetime that gives the asymptotic SdS mirror trajectory? A brief calculation gives an approximate metric of the form of Eq.~(\ref{SS}), with
\be
    f(r)\approx\frac{A(r-r_B)(r_C-r)}{r+B}.\label{f_SdSa}
\ee
Here the horizons locations $r=r_B$ and $r=r_C$ are obtained from the SdS metric, Eq.~\eqref{f_SdS1}, with $A>0$ and $B>0$ being two constants determined by imposing that the corresponding surface gravities obtained from the SdS metric, Eq.~(\ref{f_SdS1}), are the same as those obtained from the approximate metric, Eq.~(\ref{f_SdSa}). 

Additionally, we can be more precise about the all-time particle evolution in the corresponding exact spacetime. This is a tractable problem which can be answered by solving for the exact spacetime analog for Eq.~(\ref{SdSma}). The accelerated boundary correspondence (ABC) metric has a simplified spherically symmetric form of Eq.~(\ref{SS}), with
\be
    f(r)\equiv1-\frac{2M}{r}-\frac{r}{L},\label{f_ABC}
\ee
where $M>0$ and $L>0$ are two constants. We keep this notation for the constants to contrast with the SdS curved spacetime system, Eq.~(\ref{f_SdS1}). Despite the resemblance between Eqs.~(\ref{f_SdS1}) and (\ref{f_ABC}), the constants $M$ and $L$ of Eq.~(\ref{f_ABC}) do not have an immediately clear physical meaning, unlike the SdS metric. The corresponding horizons are, respectively, at $r=r_B$ and $r=r_C$ with
\be
    r_B=\frac{1}{2}\left[L-\sqrt{L(L-8M)}\right],
\ee
and
\be
    r_C=\frac{1}{2}\left[L+\sqrt{L(L-8M)}\right],
\ee
where $r_B\leq r_C$. In this case there exists no negative, nonphysical solution. This is the key point to eventually get an exact mirror with trajectory of the form of Eq.~(\ref{SdSma}). The surface gravities associated with each horizon are, respectively,
\be
    \kappa_B=\frac{r_C-r_B}{2Lr_B},
\ee
and
\be
    \kappa_C=\frac{r_C-r_B}{2Lr_C}.
\ee
If we follow the same steps as in Secs. \ref{Sec2} and \ref{Sec3} we arrived at exactly the mirror trajectory of Eq.~(\ref{SdSma}) with analog horizons $r_i$ and surface gravities $\kappa_i$ obtained from the ABC metric. In this case, the particle production results in Sec. \ref{Sec5} are completely valid at all times. Note that the SdS mirror trajectory, Eq.~(\ref{SdSm2}), has analog horizons $r_i$ and surface gravities $\kappa_i$ obtained from the SdS metric, where the particle production results in Sec. \ref{Sec5} are completely valid at early and late times.

We conclude by noting that the model in \cite{Good:2021hld} likewise presents an accelerated mirror trajectory with two different temperatures (Planck distributions) at asymptotically early and late times. This previously studied mirror has early-time and late-time thermal particle spectra, just like the SdS system. However, in \cite{Good:2021hld} there is no horizon at early times, whereas the SdS moving mirror has a cosmological horizon in this region.
 
\section{Final Remarks}\label{Sec7}
In this work, we have analyzed the horizon radiation of SdS spacetime with respect to static observers by using the moving mirror model. In particular, there a few salient steps forward made:
\begin{itemize}
    \item We have derived the accelerated boundary correspondence, Eq.~(\ref{SdSm1}) or Eq.~(\ref{SdSm2}), to SdS spacetime and determined the energy flux of the SdS moving mirror, Eq.~(\ref{EF}).
    \item We have derived the particle spectrum, Eq.~(\ref{N}), of the asymptotic SdS moving mirror, Eq.~(\ref{SdSma}).
    \item We have determined the corresponding curvature analog to the all-time particle spectrum of the asymptotic SdS moving mirror, Eq.~(\ref{SdSma}).
\end{itemize}

The temperature and thermodynamics of the SdS spacetime has been the object of study because a static observer in this universe will have to interact with both the black hole horizon and the cosmological horizon and their quantum particle creation effects. The open question of what actually constitutes a thermodynamic equilibrium in this scenario can be ignored with a small cosmological constant but becomes important with a large cosmological constant. While many previous studies have defined effective temperatures for the SdS spacetime and numerous proposals for temperature have been introduced (see e.g. \cite{Pappas:2017kam, Robson:2019yzx, Singha:2021dxe} and references therein), our work conclusively delivers on the non-thermal distribution identified with the interacting spectra. The all-time non-thermal spectrum contains definitive early-time and late-time regimes of energy flux in thermal equilibrium indicative of the particle production in different Planck distributions. It highlights the limited utility of associating an interactive temperature to describe the SdS spacetime as a whole (in agreement with \cite{Choudhury:2004ph}). While the temperature of the black hole horizon defined in terms of the corresponding surface gravity fails to take into account the lack of an asymptotically flat limit, the well-defined and expected Planck distributed particle radiation spectrum is nevertheless present in the analog moving mirror system, which is correspondingly asymptotically flat. 

\subsection*{Future Work}
While we have addressed the study of the analog SdS moving mirror and concretely tackled the problem of particle production using standard tools of quantum field theory in curved spacetime, i.e. the formalism of the Bogoliubov transformations \cite{Birrell1982}, there are future extensions. A physically appropriate approximation, i.e. the asymptotic SdS moving mirror, has been introduced and so we have managed to obtain the beta Bogoliubov coefficient, but the next step is to try to arrive at particular, intermediate-time, physical results without using any approximation. With this aim, alternative approaches that avoid the calculation of the Bogoliubov coefficients can be considered, and so we point out the formalism introduced in \cite{Fabbri:2004yy} which takes advantage of the conformal symmetry present in (1+1)-dimensional systems (like the mirror model). This formalism has been applied in black hole and cosmological scenarios \cite{Agullo:2006um, Agullo:2008qb, Agullo:2009vq, Agullo:2010hi} and it has also been applied recently to analyze the particle creation by wormholes \cite{Gurrea-Ysasi:2020mhc}. It would be insightful to try to use it in gravitational analog mirror models like the SdS ABC.

Alternatively, the extension to general cosmological black hole spacetimes would be interesting as well. In particular, the generalization to Kerr-de Sitter spacetime \cite{Bhattacharya:2018ltm, Gregory:2021ozs} could provide insight into the radiation of a rapidly spinning black hole and its evaporation process in an expanding universe. Our approach would confirm the cosmological constant quenching effect on the amplification of quantum vacuum fluctuations, among other things. 

The analogy between black holes and moving mirrors has encouraged recent fruitful investigations (see e.g. \cite{Akal:2020twv, REYES}). In this year (2021) alone, we have seen special attention given to the moving mirror model with treatments ranging from various fields and topics like entanglement capacity \cite{Kawabata:2021hac}, harvesting \cite{Liu}, holography \cite{Akal:2021foz}, qubits \cite{Agusti:2020zro}, relativistic semi-transparency \cite{Lin:2021bpe}, Page curves \cite{Ageev:2021ipd} and complexity \cite{Sato:2021ftf}. We expect the model to continue to be enriched by diverse and novel ideas connecting the nature of quantum vacuum radiation to the resulting effects of acceleration and gravity.

\acknowledgments
Thanks in memory to the late Thanu Padmanabhan. We are grateful to Alessandro Fabbri for his comments on an early version of this paper. DFS is appreciative for continuous discussions and support, and also thanks Gonzalo J. Olmo for his observations on possible future work. MG thanks Aizhan Myrzakul for discussions.

Funding from the state-targeted program “Center of Excellence for Fundamental and Applied Physics” (BR05236454) by the Ministry of Education and Science of the Republic of Kazakhstan is acknowledged, as well as the FY2021-SGP-1-STMM Faculty Development Competitive Research Grant No. 021220FD3951 at Nazarbayev University. JF acknowledges support from the Australian Research Council Centre of Excellence for Quantum Computation and Communication Technology (Project No. CE170100012).

\appendix
\section{Bogoliubov coefficient for the approximated asymptotic SdS moving mirror}\label{A}
In this Appendix, we determine the expression of the beta Bogoliubov coefficient in Eq.~(\ref{betaSdSma1}). We begin by considering Eq.~(\ref{beta}) and the asymptotic SdS mirror trajectory, Eq.~(\ref{SdSma}), to end up with
\be
    \begin{split}
        \beta_{\omega\omega'}=&\frac{1}{2\pi}\sqrt{\frac{\omega'}{\omega}}\int_{v_C}^{v_B}dv\\
        \cdot&\left[\Bar{\kappa}_B(v_B-v)\right]^{\frac{i\omega}{\kappa_B}}\left[\Bar{\kappa}_C(v-v_C)\right]^{-\frac{i\omega}{\kappa_C}}e^{-i\omega'v}.
    \end{split}
\ee
A change of variable $x=v_B-v$ allows us to arrive at
\be
    \begin{split}
        \beta_{\omega\omega'}=&-\frac{1}{2\pi}\sqrt{\frac{\omega'}{\omega}}e^{-i\omega'v_B}\int_{v_B-v_C}^0dx\\
        \cdot&\left[\Bar{\kappa}_B(x)\right]^{\frac{i\omega}{\kappa_B}}\left[\Bar{\kappa}_C(v_B-v_C-x)\right]^{-\frac{i\omega}{\kappa_C}}e^{i\omega'x}.
    \end{split}
\ee
Another change of variable $x'=\kappa'x$ with $\kappa'=(v_B-v_C)^{-1}$ leads us, after some steps, to
\be
    \begin{split}
        \beta_{\omega\omega'}=&\frac{1}{2\pi}\sqrt{\frac{\omega'}{\omega}}\frac{(\Bar{\kappa}_B)^b(\Bar{\kappa}_C)^{-c}}{(\kappa')^{1+b-c}}e^{-i\omega'v_B}\\
        \cdot&\int_0^1dx'\,(x')^b(1-x')^{-c}e^{ax'},\label{I}
    \end{split}
\ee
where we have defined
\be
    a\equiv\frac{i\omega'}{\kappa'}\quad,\quad b\equiv\frac{i\omega}{\kappa_B},\quad,\quad c\equiv\frac{i\omega}{\kappa_C}\quad.
\ee
Finally, the integral in Eq.~(\ref{I}) is analytically solved in terms of the Gamma function $\Gamma(x)$ and the confluent hypergeometric function of the first kind ${}_1F_1(p;q;z)$ as
\be
    \begin{split}
        &\int_0^1dx'\,(x')^b(1-x')^{-c}e^{ax'}=\\
        &\frac{\Gamma(1+b)\Gamma(1-c)}{\Gamma(2+b-c)}{}_1F_1(1+b;2+b-c;a),
    \end{split}
\ee
and we get
\be
    \begin{split}
      \beta_{\omega\omega'}=&\frac{1}{2\pi}\sqrt{\frac{\omega'}{\omega}}\frac{(\Bar{\kappa}_B)^b(\Bar{\kappa}_C)^{-c}}{(\kappa')^{1+b-c}}e^{-i\omega'v_B}\\
      &\cdot\frac{\Gamma(1+b)\Gamma(1-c)}{\Gamma(2+b-c)}{}_1F_1(1+b;2+b-c;a),
    \end{split}
\ee
which is the beta Bogoliubov coefficient in Eq.~(\ref{betaSdSma1}).

\bibliography{main} 

\end{document}